\documentclass[pra,twocolumn,showpacs,a4paper]{revtex4}
\usepackage{bm,graphicx,amsmath}

\newcommand{\ket}[1]{\left|#1\right\rangle}

\newcommand{\melc}[3]{\left\langle#1\left|#2\right|#3\right\rangle}

\begin{document}

\title{Assessment of a quantum phase gate operation based on nonlinear optics}
\author{S. Rebi\'{c},}
\email{stojan.rebic@unicam.it}
\author{C. Ottaviani, G. Di Giuseppe, D. Vitali and P. Tombesi}
\affiliation{Dipartimento di Fisica, Universit\`{a} di Camerino, I-62032 Camerino
(MC), Italy}

\begin{abstract}
We analyze in detail the proposal for a two-qubit gate for travelling single-photon qubits recently presented by C. Ottaviani \emph{et al}. [Phys. Rev. A \textbf{73}, 010301(R) (2006)]. The scheme is based on an ensemble of five-level atoms coupled to two quantum and two classical light fields. The two quantum fields undergo cross-phase modulation induced by electromagnetically induced transparency. The performance of this two-qubit quantum phase gate for travelling single-photon qubits is thoroughly examined in the steady-state and transient regimes, by means of a full quantum treatment of the system dynamics. In the steady-state regime, we find a general trade-off between the size of the conditional phase shift and the fidelity of the gate operation. However, this trade-off can be bypassed in the transient regime, where a satisfactory gate operation is found to be possible, significantly reducing the gate operation time.
\end{abstract}

\maketitle

\section{Introduction} \label{sec:intro}

A promising system for quantum computation is to use single photons to encode quantum information~\cite{chuang95}. This is due to the photon's robustness against decoherence and the availability of single-qubit operations. However, it is difficult to realize the necessary two-qubit operations since the physical interaction between photons is very small. Linear optics quantum computation~\cite{klm} and nonlinear optical processes involving few photons have been proposed to circumvent this problem. The first is a probabilistic scheme implicitly based on the nonlinearity hidden in single-photon detectors, while the second is based on the enhancement of photon-photon interaction either in cavity QED configurations~\cite{turch} or in dense atomic media exhibiting electromagnetically induced transparency (EIT)~\cite{eit}. In this latter case, optical nonlinearities can be produced when EIT is disturbed, either by introducing additional energy level(s)~\cite{Schmidt96,Wang01}, or by mismatching the probe and control field frequencies~\cite{Grangier98,Matsko03}.

The scope of this paper is to assess the performance of a two-qubit quantum phase gate (QPG)
for travelling single photon qubits~\cite{Lukin00,Ottaviani03,Petrosyan02,tripod,Masalas04},
based on the cross-Kerr nonlinearity which is generated in a five-level atomic medium. In a QPG, one qubit gets a phase
conditional to the other qubit state according to the transformation~\cite{Lloyd95,NielsenChuang} $|i\rangle _{1}|j\rangle _{2} \rightarrow
\exp\left\{i \phi_{ij} \right\}|i\rangle _{1}|j\rangle _{2} $ where $\{i,j\}=0,1$ denote the logical qubit bases. This gate is universal when the
conditional phase shift (CPS)
\begin{equation}\label{eq:def_cps}
\phi = \phi_{11} + \phi_{00} -\phi_{10} -\phi_{01},
\end{equation}
is nonzero, and it is equivalent to a CNOT gate up to local unitary transformations when $\phi=\pi$ \cite{Lloyd95,NielsenChuang}.
Most of the literature focused only on the evaluation of the CPS and on the best conditions for achieving
$\phi=\pi$~\cite{Lukin00,Ottaviani03,Petrosyan02,tripod,Masalas04}, while the gate fidelity, which is the main quantity
for estimating the efficiency of a gate, has been evaluated in the full quantum limit in Ref.~\cite{rapcomm} for the first time.
Here we provide the details of the calculation of the fidelity and the CPS of Ref.~\cite{rapcomm},
which showed the presence of a general \emph{trade--off} between a large CPS and a gate fidelity close to one, hindering the QPG operation,
in the stationary state. However, we shall see that this trade-off can be partially bypassed in the transient regime,
which has never been considered before in EIT situations, still allowing a satisfactory gate performance.

\begin{figure}[t]
\includegraphics[scale=0.8]{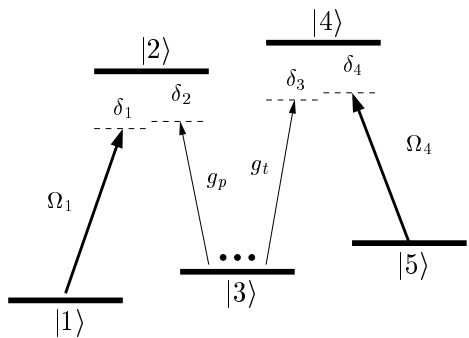}
    \caption{Energy levels of the \textit{M}-scheme. $\Omega_j$ are the Rabi frequencies of classical fields, while $g_{p,t}$
    denote couplings of the quantized probe and trigger fields to their respective transitions. $\delta_j$ are the detuning
    of the fields from resonance.} \label{fig:5levelscheme}
\end{figure}

The qubits are given by polarized single-photon wave packets with different frequencies, and the phase shifts $\phi_{ij}$ are generated
when these two pulses cross an atomic ensemble in a five-level \textit{M} configuration (see  Fig.~\ref{fig:5levelscheme}). The population is
assumed to be initially in the ground state $\ket{3}$. From this ground state, it could be excited by either the single-photon \emph{probe}
field, with central frequency $\omega_p$ and coupling to transition $\ket{3} \leftrightarrow \ket{2}$, or by the single-photon \emph{trigger} field,
with central frequency $\omega_t$ and coupling to transition $\ket{3}
\leftrightarrow \ket{4}$. We assume that the five levels are Zeeman sub-levels of an alkali atom, and that both pulses have a sufficiently narrow bandwidth.
In this way, the Zeeman splittings can be chosen so that the atomic medium is coupled only to a given circular polarization of either the probe or trigger field,
while it is transparent for the orthogonally polarized mode, which crosses the gas undisturbed~\cite{Ottaviani03}. As a consequence, the logical
basis for each qubit practically coincides with the two lowest Fock states of the mode with the ``right'' polarization, $|0_j\rangle$ and $|1_j\rangle$ ($j=p,t$),
while the ``wrong'' polarization modes will be neglected from now on.

A classical pump field, with frequency $\omega_1$ and Rabi frequency $\Omega_1$, couples to the transition $\ket{1} \leftrightarrow \ket{2}$,
while a second classical pump field, with frequency $\omega_4$
and Rabi frequency $\Omega_4$, couples to the transition $\ket{4} \leftrightarrow \ket{5}$ (see Fig.~\ref{fig:5levelscheme}).
We consider a cylindrical, quasi-1D, atomic medium with the two classical pump beams propagating
along its axis, collinear with the two quantum fields in order to avoid Doppler broadening.
When the probe field is on two-photon resonance with the pump field with Rabi frequency $\Omega_1$,
and the trigger field is on two-photon resonance with the pump field with Rabi frequency $\Omega_4$,
the system exhibits EIT for probe and trigger simultaneously.
This simultaneous EIT condition is achieved when
\begin{equation} \label{eq:twophotres}
\delta_1=\delta_2, \;\;\;\;\;\;\;\;\;\; \delta_3=\delta_4,
\end{equation}
where the detunings $\delta_j$ are defined by
\begin{subequations}
\label{eq:det2}
\begin{eqnarray}
E_{2}-E_{1}&=&\hbar\omega_{1}+\hbar\delta_{1},\\
E_{2}-E_{3}&=&\hbar\omega_{p}+\hbar\delta_{2}, \\
E_{4}-E_{3}&=&\hbar\omega_{t}+\hbar\delta_{3}, \\
E_{4}-E_{5}&=&\hbar\omega_{4}+\hbar\delta_{4}.
\end{eqnarray}
\end{subequations}
A nonzero CPS occurs whenever a nonlinear cross-phase modulation (XPM) between probe
and trigger is present. This cross-Kerr interaction takes place if the two-photon resonance condition is violated. For
small frequency mismatch $\epsilon_{12}=\delta_1-\delta_2$ and $\epsilon_{34}=\delta_3-\delta_4$ (both chosen to be within the
EIT window), absorption remains negligible and the cross-Kerr interaction between probe and trigger photons may be strong.
The consequent CPS may become large, of the order of $\pi$, if the probe and trigger pulse interact for a sufficiently long time.
If the two single photon pulses enter simultaneously the
atomic medium, their interaction time $t_{int}$ is optimized when the group velocities of the two pulses are equal, so that
$t_{int}=L/v_g$, where $v_g$ is the common group velocity of the pulses and $L$ is the length of the gas cell.
The inherent \emph{symmetry} of the scheme guarantees perfect group velocity matching for probe and trigger whenever $\delta_1=\delta_4$, $\delta_2=\delta_3$
and $g_{p}/\Omega_{1}=g_{t}/\Omega_{4}$, where $g_j=\mu_j \sqrt{\omega_j/2\hbar \epsilon_0 V_j}$ ($j=p,t$) is the coupling constant
between the quantum mode with frequency $\omega_j$ and mode volume $V_j$, and the corresponding transition with electric dipole moment $\mu_j$.

The importance of group velocity matching for achieving a significant nonlinear phase shift
was first pointed out in Ref.~\cite{Lukin00}, which suggested to use a mixture of two different atomic species to achieve this goal.
The first kind of atoms generates XPM by means of a four-level \emph{N} scheme \cite{Schmidt96}, in which however
only the probe field undergoes EIT and it is slowed down,
while the second kind of atoms realizes a three-level $\Lambda$ scheme able to slow down the trigger
pulse. Group velocity matching is achieved by means of an accurate but difficult control of the atomic densities.
A different way of achieving group velocity matching, but which is still asymmetric for probe and trigger
has been proposed in Ref.~\cite{Ottaviani03}, which employs a five-level \emph{M} scheme similar to the one discussed here,
but with a different atomic population distribution. In that case, the two group velocities can be tuned and made
equal simply by tuning the frequencies and intensities of the two classical pump fields.
Instead, Ref.~\cite{Petrosyan02} considered a six level scheme in which probe and trigger are affected
by EIT and XPM in a symmetric fashion, so that the corresponding
group velocities are equal by construction. The present proposal achieves group velocity matching just in the same way
(see also Refs.~\cite{tripod}, where a four-level tripod system, symmetric between probe and trigger, has been
proposed for XPM).

The paper is organized as follows. In Sec.~\ref{sec:model} we describe the model used in the remainder of the paper.
Sec.~\ref{sec:perturbative} shows the results of a perturbative calculation for the CPS. These are used as a
motivation to pass to a density matrix based calculations in Sec.~\ref{sec:SSOperation}, describing a QPG operation in a steady-state.
Then, the transient regime in explored in Sec.~\ref{sec:transient}, while in Sec.~\ref{sec:exper} a scheme for the
experimental verification of the QPG operation is discussed in detail. Conclusions are given in Sec.~\ref{sec:conclusion}.

\section{Model} \label{sec:model}

\begin{figure}[t]
\includegraphics[scale=0.8]{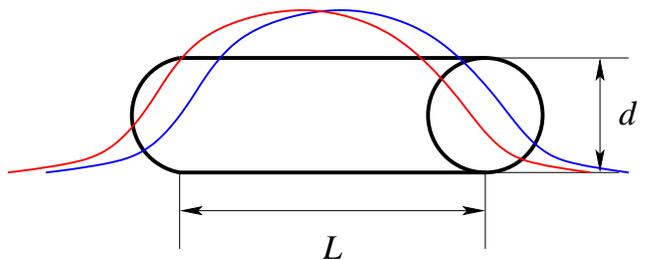}
    \caption{(Color online) A schematic plot of the assumed single-photon pulse propagation through the gas cell of length $L$ and diameter $d$.
    The pulse length is assumed to coincide with the cell length $L$, and the pulse waist $w$ is assumed to be of order of cell diameter $d$.} \label{fig:pulsescheme}
\end{figure}

In this Section, we present the model we have adopted for a full quantum description of the interaction of the two single-photon wave-packets
with the atomic medium possessing the level structure outlined in Fig.~\ref{fig:5levelscheme}.
To this end, we make the following two assumptions which, even though not simple to realize experimentally, are more technical than physical in nature:
\begin{enumerate}
\item We assume perfect spatial mode matching between the input single-photon pulses entering the gas cell and the optical modes
naturally excited by the driven atomic medium, and which are determined by the geometrical configuration of the gas cell and of the pump beams \cite{Duan02}.
This allows us to describe the probe and trigger fields with the right polarization
in terms of \emph{single} travelling optical modes, with annihilation operators $\hat{a}_{p,t}$.
\item We assume that the pulses are tailored in such a way that they simultaneously enter gas cell and completely overlap with it
during the interaction (see Fig.~\ref{fig:pulsescheme}). This means that their length (compressed inside a medium due to group
velocity reduction) is of the order of the cell length $L$ and their beam waist is of the order of the cell radius. In this way,
the two pulses interact with \emph{all} $N_a$ atoms in the cell at once, and moreover, one can ignore spatial aspects of pulse propagation.
\end{enumerate}
With these assumptions, and neglecting dipole-dipole interactions, the interaction picture Hamiltonian may be written as
\begin{eqnarray}\label{eq:coll_ham}
&& H = \hbar\epsilon_{12}\hat{S}_{11}+\hbar\delta_{2}\hat{S}_{22} +\hbar\delta_{3}\hat{S}_{44}+\hbar\epsilon_{34}\hat{S}_{55}  \\
&&+\hbar\Omega_{1}\sqrt{N_a}\left(\hat{S}_{21}+\hat{S}_{12}\right)+\hbar g_{p}\sqrt{N_{a}}\left(\hat{a}_{p}\hat{S}_{23}+\hat{S}_{32}\hat{a}_{p}^{\dagger}\right)
 \nonumber\\
&&+\hbar g_{t} \sqrt{N_{a}} \left(\hat{a}_{t}\hat{S}_{43}+\hat{S}_{34}\hat{a}_{t}^{\dagger}\right)+ \hbar\Omega_{4}\sqrt{N_a} \left(\hat{S}_{45}+\hat{S}_{54}\right),
\nonumber
\end{eqnarray}
where we have defined the collective atomic operators
\begin{subequations}
\label{eq:collectiveop}
\begin{eqnarray}\label{eq:collective1}
\hat{S}_{kl}&=&\frac{1}{\sqrt{N_{a}}}\sum_{i=1}^{N_a}\sigma_{kl}^i, \;\;\;\;k\neq l=1,\ldots,5, \\
\hat{S}_{kk}&=& \sum_i^{N_a}\sigma_{kk}^i, \label{eq:collective2}
\end{eqnarray}
\end{subequations}
with $\sigma_{kl}^i\equiv |k\rangle _i \langle l |$ being the operator switching between
states $k$ and $l$ of the $i$th atom. The initial state of the system corresponds to a probe and a trigger single-photon pulse with generic polarization,
simultaneously entering the medium in which all the atoms are initially in state $|3\rangle$. Since we consider only the polarization mode interacting with the medium,
both for the probe and the trigger, the initial state can be written as
\begin{eqnarray}
|\psi_{in}\rangle &=& \bigotimes_{i=1}^{N_a} \ket{3}_i \otimes \left(c_{00}|0_p\rangle \otimes |0_t\rangle+c_{01}|0_p\rangle \otimes|1_t\rangle \right.
\nonumber \\
&& \left. +c_{10}|1_p\rangle \otimes |0_t\rangle
+c_{11}|1_p\rangle \otimes |1_t\rangle \right). \label{iniz}
\end{eqnarray}
Due to the above assumptions, the passage of the two pulses through the atomic medium of length $L$ corresponds to the time evolution of this state, for a time
$t_{int}=L/v_g$, according to the master equation \cite{FleischhauerRMP}
\begin{eqnarray} \label{eq:mastercoll}
\dot{\rho} &=& -\frac{i}{\hbar}\left[ H,\, \rho\right] +\sum_{k}\frac{\gamma_{kk}}{2}\sum_{j=1}^{N_a}
\left( 2\sigma_{kk}^j\rho\sigma_{kk}^j - \sigma_{kk}^j\rho
- \rho\sigma_{kk}^j \right),\nonumber \\
&& + \sum_{kl}\frac{\gamma_{kl}}{2}\sum_{j=1}^{N_a}\left( 2\sigma_{kl}^j\rho\sigma_{kl}^{j \dagger}
- \sigma_{kl}^{j \dagger} \sigma_{kl}^{j}\rho
- \rho\sigma_{kl}^{j \dagger} \sigma_{kl}^{j}\right),
\end{eqnarray}
including not only the coherent interaction described by the Hamiltonian of Eq.~(\ref{eq:coll_ham}), but also the spontaneous emission
from the excited states $l=2,4$ to the ground states $k=1,3,5$ ($\gamma_{kl}$ denotes the corresponding decay rate)
and the dephasing of levels $|k\rangle $, $k=1,2,4,5$, with dephasing rate $\gamma_{kk}$. Typically the dephasing rates are much smaller
than the decay rates, $\gamma_{kl} \gg \gamma_{kk}$, $\forall \;k,l$.

Since the initial state of Eq.~(\ref{iniz}) contains at most two excitations, the coherent time evolution driven by Eq.~(\ref{eq:coll_ham})
is simple and restricted to a finite-dimensional Hilbert
space involving few symmetric collective atomic states. In fact, each component of the initial state of Eq.~(\ref{iniz})
evolves independently in a different subspace. The component with no photon is an eigenstate of $H$ and does not evolve.
The $\bigotimes_{i=1}^{N_a} \ket{3}_i |0_p\rangle \otimes  |1_t\rangle$ component evolves in a three-dimensional Hilbert space which it
spans together with the two states $|e_4^{(0,0)}\rangle $ and $|e_5^{(0,0)}\rangle$. Here, we have defined the symmetric collective states
\begin{equation}\label{eq:gen_state}
|e_r^{(n_p,n_t)}\rangle  = \frac{1}{\sqrt{N_{a}}}\sum_{i=1}^{N_{a}}\ket{3_{1},3_{2},\dots,r_{i},\dots,3_{N_{a}}} \otimes\ket{n_{p}}\otimes\ket{n_{t}},
\end{equation}
where $r = 1,2,4,5$. In a similar fashion, the component with only one probe photon evolves in a three-dimensional Hilbert
space spanned by the three states $\bigotimes_{i=1}^{N_a} \ket{3}_i |1_p\rangle \otimes  |0_t\rangle$, $|e_1^{(0,0)}\rangle $ and $|e_2^{(0,0)}\rangle $.
The component with one probe and one trigger photon evolves in the five dimensional subspace spanned by the four collective states $|e_1^{(0,1)}\rangle $,
$|e_2^{(0,1)}\rangle $, $|e_4^{(1,0)}\rangle $, and $|e_5^{(1,0)}\rangle$ and the state $\bigotimes_{i=1}^{N_a} \ket{3}_i |1_p\rangle \otimes  |1_t\rangle$.

Decoherence effects, and more specifically spontaneous emission from each atom complicates this dynamics.
However, we are in the weak excitation limit where, for $l \neq 3$, $\langle \sigma_{ll}^j\rangle \simeq N_a^{-1} \ll 1$,
as shown by the fact that the Hamiltonian dynamics involve only the symmetric atomic
states of the form of Eq.~(\ref{eq:gen_state}). This limit allows a drastic simplification of the effective time evolution.
Following Duan {\em et al.}~\cite{Cirac02}, we can introduce Fourier transforms of the individual atomic operators
$s^\mu_{kl} = \sum_{j=0}^{N_a-1}\sigma_{kl}^je^{ij\mu/N_a}/\sqrt{N_a}$, where $s^0_{kl}=\hat{S}_{kl}$ are the collective operators defined in
Eq.~(\ref{eq:collective1}). The sum over the atoms in Eq.~(\ref{eq:mastercoll}) then transforms to the sum over the collective atomic modes with index
$\mu$. In the weak excitation limit,
the operators $s^\mu_{kl}$ approximately commute with each other. This means that they represent independent collective atomic modes,
and one can trace over the $\mu \neq 0$ modes, so that the spontaneous emission term in the master equation becomes
\begin{equation}
\sum_{kl}\frac{\gamma_{kl}}{2}\left( 2\hat{S}_{kl}\rho\hat{S}_{kl}^\dagger
- \hat{S}_{kl}^\dagger\hat{S}_{kl}\rho - \rho\hat{S}_{kl}^\dagger\hat{S}_{kl} \right),
\label{eq:mastereff}
\end{equation}
where the sum is now over the six ``collective'' spontaneous decay channels only, each characterized by the \emph{single-atom}
decay rate $\gamma_{kl}$. A similar argument applies to the dephasing term in the master equation (\ref{eq:mastercoll}). In fact,
if we restrict to the subspace of the symmetric collective states of Eq.~(\ref{eq:gen_state}) involving only
single atomic excitations, we can approximate in the dephasing terms of the master
equation,
\begin{equation} \label{eq:mastereff2}
\sum_{k}\gamma_{kk}\sum_{j=1}^{N_a}
\sigma_{kk}^j\rho\sigma_{kk}^j \simeq \sum_{k}\gamma_{kk} \hat{S}_{kk}\rho\hat{S}_{kk},
\end{equation}
where $ \hat{S}_{kk}$ is given by Eq.~(\ref{eq:collective2}). Using Eqs.~(\ref{eq:mastereff}) and (\ref{eq:mastereff2}), the
master equation of Eq.~(\ref{eq:mastercoll}) in the weak excitation limit becomes
\begin{eqnarray} \label{eq:mastercoll2}
\dot{\rho} &=& -\frac{i}{\hbar}\left[ H,\, \rho\right] +\sum_{k}\frac{\gamma_{kk}}{2} \left(2\hat{S}_{kk}\rho\hat{S}_{kk}-
\hat{S}_{kk}\rho - \rho\hat{S}_{kk}\right)
\nonumber \\
&& + \sum_{kl}\frac{\gamma_{kl}}{2}\left( 2\hat{S}_{kl}\rho\hat{S}_{kl}^\dagger
- \hat{S}_{kl}^\dagger\hat{S}_{kl}\rho - \rho\hat{S}_{kl}^\dagger\hat{S}_{kl} \right),
\end{eqnarray}
that is, it involves only the operators of the collective atomic mode with index $\mu=0$.
This actually means that the single photon probe and
trigger pulses excite only a restricted number
of collective atomic states, so that the atomic medium behaves as an effective \emph{single} $5$-level atom, with a \emph{collectively enhanced}
coupling with the optical modes $g_j\sqrt{N_a}$, but with single-atom decay rates $\gamma_{kl}$, dephasing rates $\gamma_{kk}$,
Rabi frequencies $\Omega_i$, and detunings $\delta_i$.

Spontaneous emission causes the four independent Hilbert subspaces corresponding to the four initial state components to become coupled.
Moreover, the ``cross'' decay channels $|4\rangle \to |1\rangle $ and $|2 \rangle \to |5\rangle $ couple the above-mentioned collective
states with six new states, $|e_1^{(1,0)}\rangle$, $|e_2^{(1,0)}\rangle $, $|e_3^{(2,0)}\rangle $ (populated if $\gamma_{41}\neq 0$),
and $|e_5^{(0,1)}\rangle $, $|e_4^{(0,1)}\rangle $,  $|e_3^{(0,2)}\rangle $ (populated if $\gamma_{25}\neq 0$). Therefore Eq.~(\ref{eq:mastercoll2})
actually describes dynamics in a Hilbert space of dimension $18$, which we have numerically solved in order to establish the performance of the QPG.
Notice that, due to the combined action of the cross-decay channels and of the Hamiltonian (\ref{eq:coll_ham}),
the states $|e_1^{(1,0)}\rangle$, $|e_2^{(1,0)}\rangle $, $|e_5^{(0,1)}\rangle $ and $|e_4^{(0,1)}\rangle $
are coupled also to \emph{doubly excited} atomic collective states without photons which are neglected by our treatment. However,
as we shall see below in the paper,
a good QPG performance is possible only when spontaneous emission events are rare. Under this condition, the probability to populate these doubly excited
atomic collective states during the atom-field interaction is completely negligible, and therefore our model based on the effective single five-level atom
description provided by Eq.~(\ref{eq:mastercoll2}) is essentially correct.

In summary, the model outlined above relies on the single-photon nature of the excitations. In this case, the collective operators~(\ref{eq:collectiveop}) effectively switch between the states making the superposition $\ket{\psi_{in}}$~(\ref{iniz}), and the symmetric collective states $\ket{e_r^{(n_p,n_t)}}$~(\ref{eq:gen_state}). This is central to the reasoning leading to the effective master equation~(\ref{eq:mastercoll2}).

To characterize the QPG operation, we calculate the CPS $\phi$ of Eq.~(\ref{eq:def_cps})
and the {\em fidelity} of the gate. The accumulated CPS $\phi$ as a function of the interaction time $t_{int}$ is obtained by using the fact
that the phase shifts $\phi_{ij}$ of Eq.~(\ref{eq:def_cps}) are given by combinations of the phases of the off-diagonal matrix elements (in the Fock basis)
of the reduced density matrix of the probe and trigger modes, $\rho_f(t) = \textrm{Tr}_{atoms}\{\rho (t) \}$.

The gate fidelity is given by \cite{NielsenChuang}
\begin{equation} \label{eq:fid}
{\mathcal F}(t) = \sqrt{\overline{\melc{\psi_{id}(t)}{\rho_{f}(t)}{\psi_{id}(t)}}},
\end{equation}
where
\begin{eqnarray} \label{eq:psi_id}
\ket{\psi_{id}(t)} &=& c_{00}\exp\{i\phi_{00}(t_{int})\} |0_p,0_t\rangle \nonumber \\ && + c_{01}\exp\{i\phi_{01}(t)\}|0_p,1_t\rangle \nonumber \\
&& + c_{10}\exp\{i\phi_{10}(t)\} |1_p,0_t\rangle \nonumber \\
&& + c_{11}\exp\{i\phi_{11}(t)\}|1_p,1_t\rangle
\end{eqnarray}
is the ideally evolved state from the initial condition (\ref{iniz}), with phases $\phi_{ij}(t)$ evaluated from $\rho_f(t)$ as discussed above.
The overbar denotes the average over all initial states (i.e., over the $c_{ij}$, see Poyatos \emph{et al.}~\cite{Poyatos97}).
The above fidelity characterizes the performance of the QPG as a deterministic gate. However, one could also consider the QPG as a
\emph{probabilistic} gate, whose operation is considered only when the number of output photons is equal to the number of input photons.
The performance of this probabilistic QPG could be experimentally studied by performing a conditional  detection of the phase shifts,
and it is characterized by the \emph{conditional} fidelity ${\mathcal F}^c(t)$, which will be discussed in Sec.~\ref{sec:SSOperation}.

\section{Perturbative Regime} \label{sec:perturbative}

The conditional fidelity is always larger than the unconditional one, but they become equal (and both approach $1$) for an ideal QPG in
which the number of photons is conserved and all the atoms remain in state $|3\rangle$. This ideal condition is verified in the limit of
large detunings $\delta_j \gg \gamma_{kj}$ so that spontaneous emission is significantly suppressed and can be neglected,
and very small couplings $g_j\sqrt{N_a}\ll \Omega_j$.
In this limit, each component of the initial state of Eq.~(\ref{iniz}) practically coincides with the dark state of the four independent Hamiltonian
dynamics discussed in Sec.~\ref{sec:model}. The system with the initial state containing zero probe and trigger photons does not evolve,
i.e. stays in the initial state $\bigotimes_{i=1}^{N_a} \ket{3}_i |0_p\rangle \otimes  |0_t\rangle$ all the time.
\begin{figure*}[t]
\begin{center}
\hfill\includegraphics[scale=0.7]{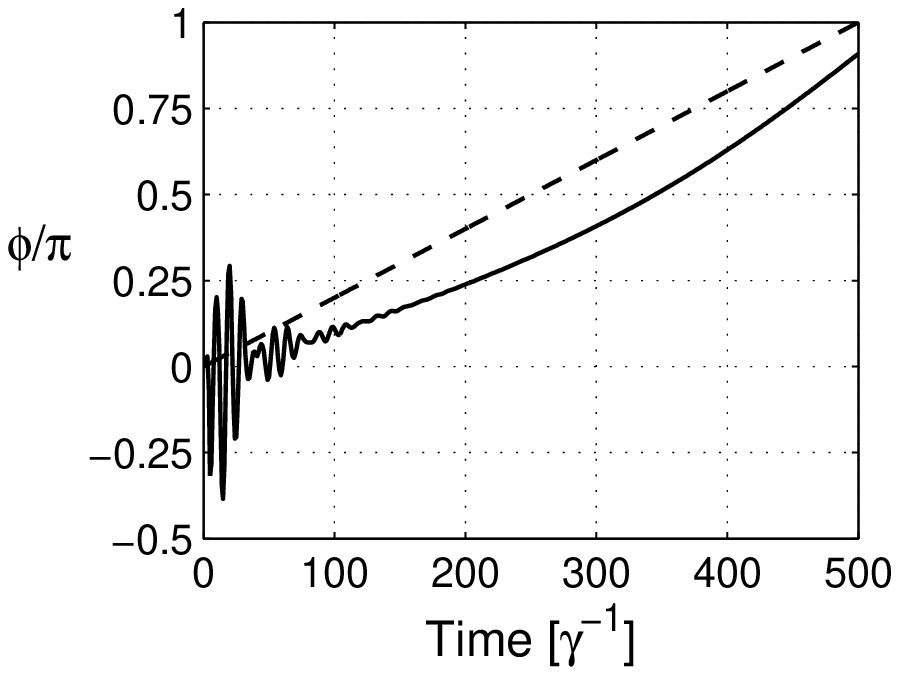}%
\hfill\includegraphics[scale=0.85]{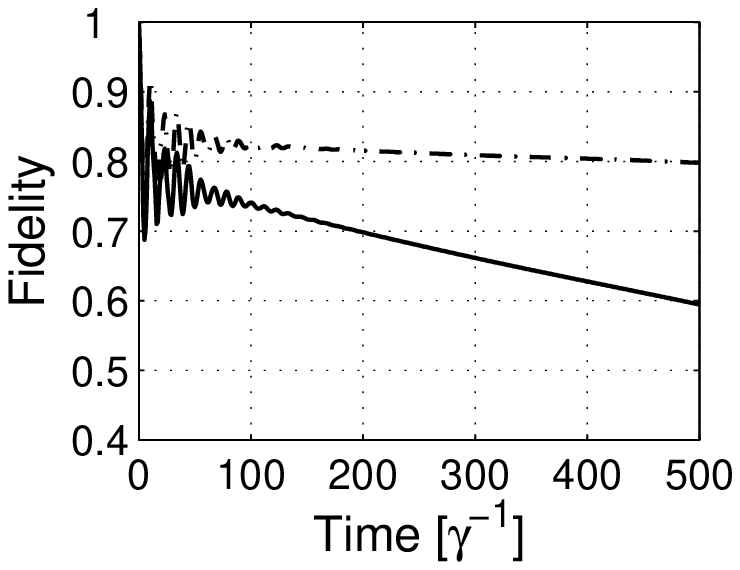}\hspace*{\fill}
\end{center}
    \caption{Conditional phase shift (\textit{left}) and the fidelity of Eq.~(\protect\ref{eq:fid}) (\textit{right}) as a function of the interaction time for $N_a=10^6$, $\delta_1=\delta_3=7.5\gamma$, $\epsilon_{12}=\epsilon_{34}=0.05\gamma$,
    $g_p=g_t=0.0011\gamma$, $\Omega_1=\Omega_4=1.875\gamma$ and $\gamma_{kk}=\gamma_{ph} = 10^{-3}\gamma$, $\forall \; k$.
    We have taken equal decay rates, $\gamma_{21}=\gamma_{23}=\gamma_{25}=\gamma_{41}=\gamma_{43}=\gamma_{45}=\gamma/3$,
    with $\gamma = 2\pi\times 6$ MHz.
    \textit{Left}: solid line represents the phase shift, as calculated from the full master equation, while the dashed line gives the perturbative prediction
    of Eq.~(\protect\ref{eq:phipert}). \textit{Right}: solid line is the unconditional gate fidelity ${\mathcal F}(t)$, while the dot-dashed line
    is the conditional one, ${\mathcal F}^c(t)$.}
\label{fig:CPSlong}
\end{figure*}
The subsystem containing one probe and zero trigger photons as the initial state evolves according to a reduced three-dimensional Hamiltonian, which
in the basis formed by the states $\bigotimes_{i=1}^{N_a} \ket{3}_i |1_p\rangle \otimes  |0_t\rangle$, $|e_2^{(0,0)}\rangle $ and $|e_1^{(0,0)}\rangle $,
is given by
\begin{subequations}
\label{eq:HamiltonianMatrices}
\begin{equation}\label{eq:Hp}
H_{p} = \left(\begin{array}{ccc}0 & g_p\sqrt{N_{a}} & 0 \\g_p\sqrt{N_{a}} & \delta_{2} & \Omega_{1} \\0 & \Omega_{1} & \epsilon_{12}\end{array}\right).
\end{equation}
Similarly, the subsystem containing one trigger and zero probe photons as the initial state evolves according to a reduced three-dimensional Hamiltonian, which
in the basis formed by the states $\bigotimes_{i=1}^{N_a} \ket{3}_i |0_p\rangle \otimes  |1_t\rangle$, $|e_4^{(0,0)}\rangle $ and $|e_5^{(0,0)}\rangle $,
is given by
\begin{equation}\label{eq:Ht}
H_{t} = \left(\begin{array}{ccc}0 & g_t\sqrt{N_{a}} & 0 \\g_t\sqrt{N_{a}} & \delta_{3} & \Omega_{4} \\0 & \Omega_{4} & \epsilon_{34}\end{array}\right).
\end{equation}
Finally, the subsystem containing initially one photon each in probe and trigger modes evolves according to a reduced five-dimensional Hamiltonian, which
in the basis formed by the states $|e_1^{(0,1)}\rangle $,
$|e_2^{(0,1)}\rangle $, $\bigotimes_{i=1}^{N_a} \ket{3}_i |1_p\rangle \otimes  |1_t\rangle$, $|e_4^{(1,0)}\rangle $, and $|e_5^{(1,0)}\rangle$,
is given by
\begin{equation}\label{eq:Hpt}
H_{pt}=\left(\begin{array}{ccccc}\delta_{2} & \Omega_{1} & g_{p}\sqrt{N_{a}} & 0 & 0 \\\Omega_{1} & \epsilon_{12} & 0 & 0 & 0 \\g_{p}\sqrt{N_{a}}
& 0 & 0 & g_{t}\sqrt{N_{a}} & 0 \\0 & 0 & 0 & \delta_{3} & \Omega_{4} \\0 & 0 & g_{t}\sqrt{N_{a}} & \Omega_{4} & \epsilon_{34}\end{array}\right).
\end{equation}
\end{subequations}
The phase accumulation experienced by the various components of the quantum state of the fields will be proportional to the eigenvalues of these matrices.
The four phase shifts $\phi_{ij}$ can be evaluated as a fourth-order perturbation expansion of the eigenvalue corresponding to the dark state in each subspace,
multiplied by the interaction time $t_{int}$.
The CPS is then calculated as
\begin{equation}\label{eq:cps_eigs}
\phi = (\lambda_{H_{pt}} - \lambda_{H_{p}} - \lambda_{H_{t}}) t_{int},
\end{equation}
where the $\lambda$'s denote the eigenvalues of the corresponding reduced Hamiltonian, with $\lambda_{H_{pt}}
\leftrightarrow \phi_{11}$, $\lambda_{H_{p}} \leftrightarrow \phi_{10}$, $\lambda_{H_{t}} \leftrightarrow \phi_{01}$
and $\phi_{00} = 0$, in agreement the with general definition of Eq.~(\ref{eq:def_cps}). Following this procedure results in the following CPS
\begin{eqnarray}\label{eq:phipert}
\phi &=& \frac{g_{p}^{2}g_{t}^{2}N_a^2 t_{int}}{(\epsilon_{34}\delta_{3} - \Omega_{4}^{2})(\epsilon_{12}\delta_{2} - \Omega_{1}^{2})}
\times \nonumber \\ &&\times\left[\frac{\epsilon_{34}(\epsilon_{12}^{2}+\Omega_{1}^{2})}{(\epsilon_{12}\delta_{2} - \Omega_{1}^{2})}
+ \frac{\epsilon_{12}(\epsilon_{34}^{2} + \Omega_{4}^{2})}{(\epsilon_{34}\delta_{3} - \Omega_{4}^{2})}\right].
\end{eqnarray}
This prediction is verified by the numerical solution of Eq.~(\ref{eq:mastercoll2})
in the limit of large detunings and small couplings. However the resulting CPS is too small,
even for very long interaction times (i.e., long gas cells): for example, for $g_{p,t}\sqrt{N_a}=0.5$ MHz,
$\epsilon_{12,34}=1.9$ MHz, $\Omega_{1,4}=65$ MHz and $\delta_{2,3}=1.9$ GHz, we obtain a tiny CPS of only
$3\times 10^{-4}$ radians when $t_{int}=10^{-4}$ s, which corresponds to $L\simeq 30$ km. This is not surprising because this limit corresponds
to a dispersive regime far from EIT. In this regime, transparency is achieved by means of a strong coupling field,
producing a well-separated Autler-Townes doublet~\cite{FleischhauerRMP}. At the same time, the size of nonlinearity is small due
to the extremely weak coupling of the quantized fields to their respective transitions. The results are, therefore, not different
from those expected from XPM in a standard optical fiber~\cite{AgarwalBook}.
Therefore, one has to explore the non-perturbative regime of larger couplings in order to exploit the low-noise,
large-nonlinearity properties of EIT and achieve a satisfactory QPG operation.
\begin{figure*}[t]
\begin{center}
\hfill\includegraphics[scale=0.7]{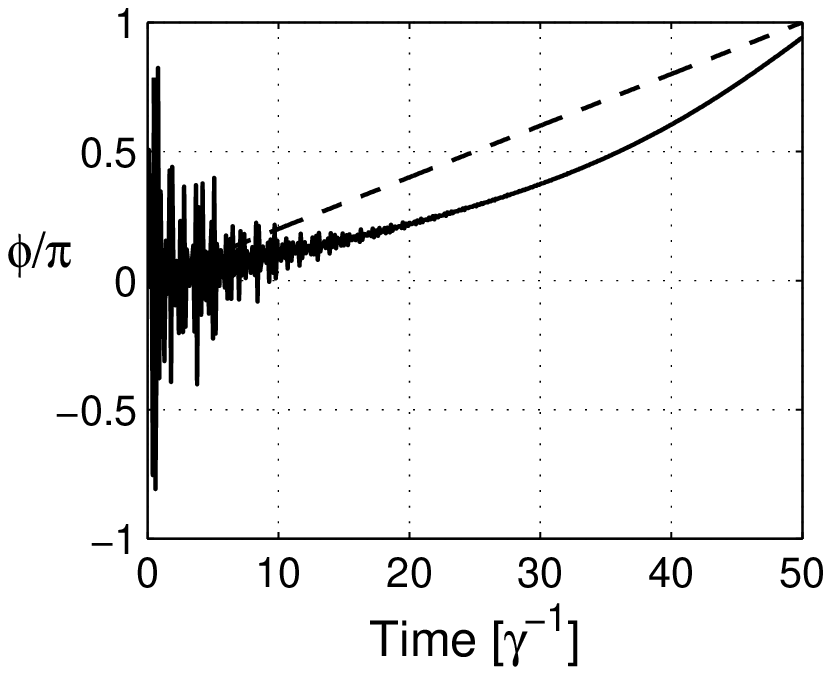}%
\hfill\includegraphics[scale=0.82]{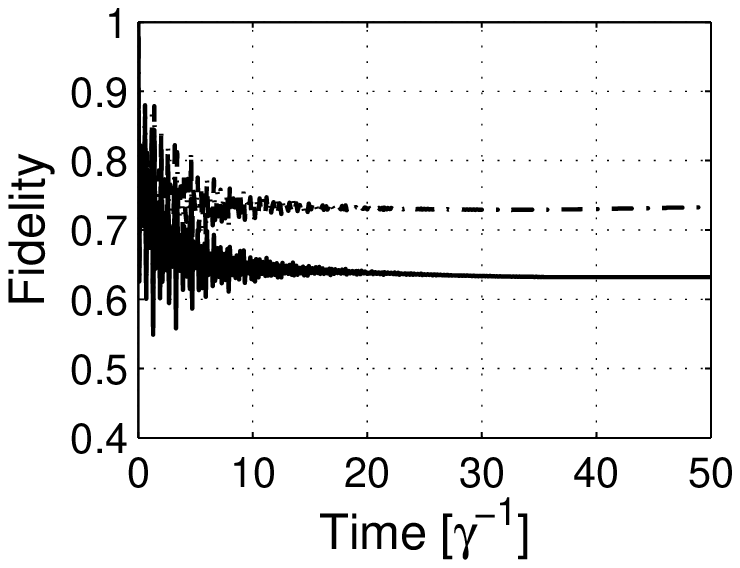}\hspace*{\fill}
\end{center}
    \caption{Conditional phase shift (\textit{left}) and the fidelity (\textit{right}) as a function of the interaction time for $N_a=10^8$, $\delta_1=\delta_3=9.5\gamma$,
    $\epsilon_{12}=\epsilon_{34}=0.2\gamma$, $g_p=g_t=0.018\gamma$ and $\Omega_1=\Omega_4=19\gamma$ and $\gamma_{kk}=\gamma_{ph} = 10^{-3}\gamma$, $\forall \; k$.
    We have taken equal decay rates, $\gamma_{21}=\gamma_{23}=\gamma_{25}=\gamma_{41}=\gamma_{43}=\gamma_{45}=\gamma/3$,
    with $\gamma = 2\pi\times 6$ MHz. \textit{Left}: solid line represents the phase shift,
    as calculated from a density matrix, while the dashed line gives the eigenvalue solution. \textit{Right}: solid line is the unconditional gate fidelity ${\mathcal F}(t)$, while the dot-dashed line is the conditional one, ${\mathcal F}^c(t)$.}
\label{fig:CPSshort}
\end{figure*}

\section{Steady-State QPG Operation} \label{sec:SSOperation}

A large amount of work exploring EIT-based nonlinear optical phenomena considers the steady-state of a generic EIT-based
system as being the natural state in which to predict and test different phenomena~\cite{Schmidt96,Ottaviani03,Matsko03,Wang01}.
We shall see in this Section that it is not possible to achieve a satisfactory QPG
performance in such a steady-state regime.

In this Section, we analyze the performance of the QPG at the steady state. To this end, we consider two different
parameter regimes: \emph{(i)} the regime of long interaction times, a natural extension of the perturbation
analysis of Sec.~\ref{sec:perturbative}, and \emph{(ii)} the regime of short interaction times, corresponding to a
non-perturbative regime with strong atom-field coupling.

\subsection{Long Interaction Time} \label{sec:SSLong}

Naturally extending the perturbative analysis, we solve the master equation~(\ref{eq:mastercoll2}), and show the results in
Fig.~\ref{fig:CPSlong}. Fig.~\ref{fig:CPSlong} (\textit{left}) shows the result for the conditional phase shift. Solid line has been calculated from the solution of Eq.~(\ref{eq:mastercoll2}), as explained in Sec.~\ref{sec:model}.
The dashed line is the `benchmark' solution, obtained from the eigenvalues of the associated Hamiltonian, by using Eq.~(\ref{eq:cps_eigs}).
The eigenvalues of the Hamiltonians of Eqs.~(\ref{eq:HamiltonianMatrices}) have been calculated numerically for the set of parameters
of Fig.~\ref{fig:CPSlong}.

It is evident that the `benchmark' solutions offers a reasonably good estimate for the size of the CPS.
The exact dynamics driven by the master equation~(\ref{eq:mastercoll2}) presents an additional
oscillatory behavior both on a short time scale (transient processes),
and on a long-time scale. The longer time scale comes from the fact that to induce the cross-Kerr nonlinearity,
one has to detune the fields away from the dark resonance. This detuning is very small and is seen in the
oscillations on a long time-scale. As both probe and trigger fields are detuned by the same amount,
only one frequency of long-time oscillations is observed.

In Fig.~\ref{fig:CPSlong} (\text{right}), fidelities (averaged over all possible two-qubit initial states) are shown in two cases.
Both are calculated by using Eq.~(\ref{eq:fid}), but they differ in the way $\rho_f(t)$ is defined. The
solid line in Fig.~\ref{fig:CPSlong} (\text{right}) refers to the \emph{unconditional} fidelity ${\mathcal F}(t)$, which is calculated from Eq.~(\ref{eq:fid})
by taking $\rho_f(t) = {\rm Tr}_{at}\{\rho(t)\}$, where $\rho(t)$ is the solution of the master equation~(\ref{eq:mastercoll2}).
The unconditional fidelity quantifies the performance of the QPG as
a \emph{deterministic} gate for single-photon qubits.

The dot-dashed line in Fig.~\ref{fig:CPSlong} (\textit{right}) refers to the \emph{conditional} fidelity
${\mathcal F}^c(t)$, which is evaluated according to Eq.~(\ref{eq:fid}), but with $\rho_f(t)$ replaced by
$\rho_f^c(t)={\rm Tr}_{at}\{|\psi_{nj}(t)\rangle \langle \psi_{nj}(t)|\}/\langle \psi_{nj}(t)|\psi_{nj}(t)\rangle $,
where $|\psi_{nj}(t)\rangle$ is the (non-normalized) evolved atom-field state conditioned to the detection of no quantum
jumps~\cite{Carmichael93}, i.e., of no photon loss by spontaneous emission. This fidelity can be measured by
post-selecting those measurement results that conserve photon number, i.e., discarding those data sets where at least a photon
from the initial two-qubit state has been lost to the environment. The conditional fidelity quantifies the performance of the QPG
as a \emph{probabilistic} two-qubit gate.

\begin{figure*}[t]
\begin{center}
\hfill\includegraphics[scale=0.7]{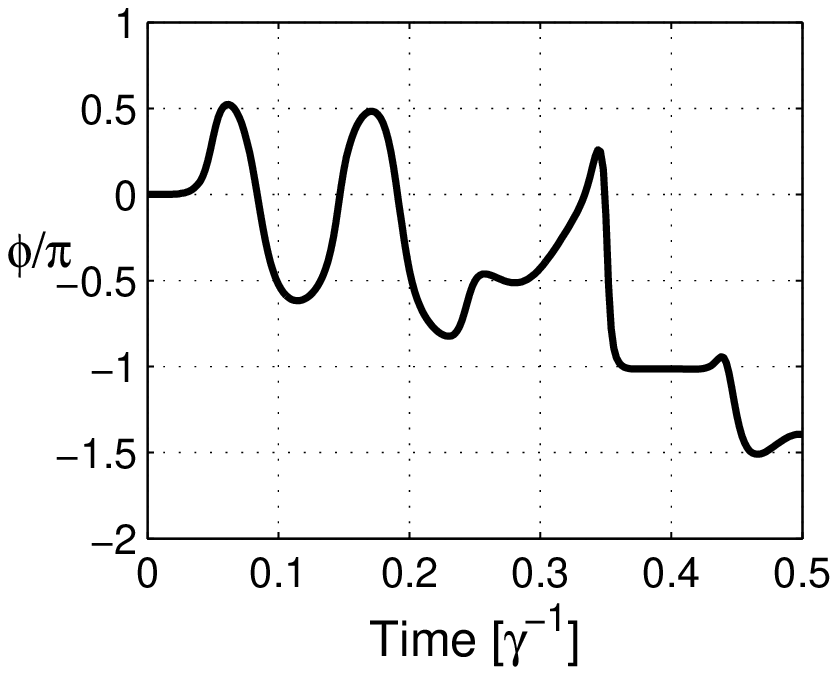}%
\hfill\includegraphics[scale=0.7]{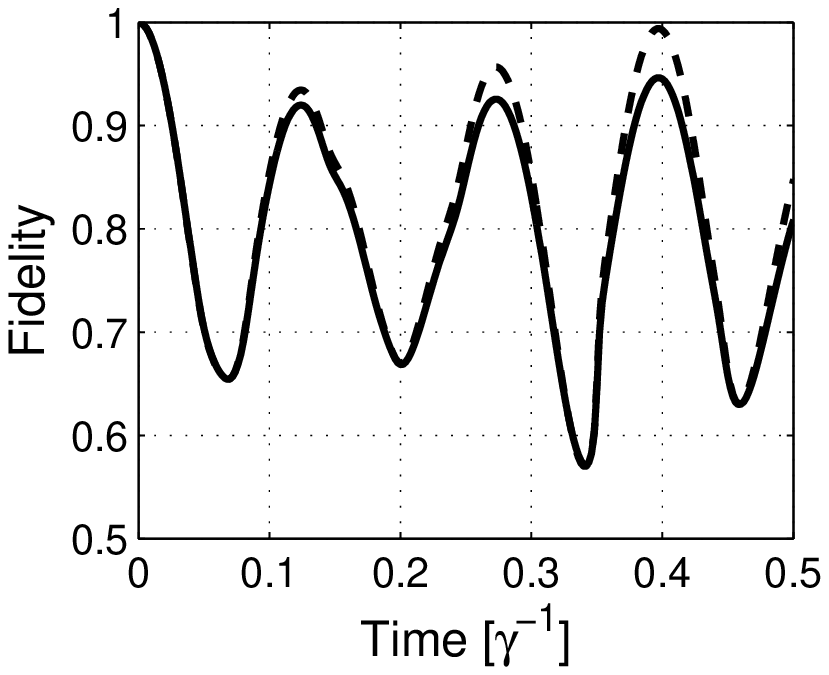}\hspace*{\fill}
\end{center}
    \caption{Conditional phase shift (\textit{left}) and the fidelity (\textit{right}) of the QPG operation for $N_a=10^6$, $\delta_1=\delta_3=15\gamma$, $\epsilon_{12}=\epsilon_{34}=0.01\gamma$, $g_p=g_t=0.0022\gamma$,
    $\Omega_1=\Omega_4=4\gamma$ and $\gamma_{kk}=\gamma_{ph} = 10^{-3}\gamma$, $\forall \; k$.
    We have taken equal decay rates, $\gamma_{21}=\gamma_{23}=\gamma_{25}=\gamma_{41}=\gamma_{43}=\gamma_{45}=\gamma/3$,
    with $\gamma = 2\pi\times 6$ MHz. \textit{Right}: the unconditional fidelity (solid) and conditional fidelity (dashed) are shown. See text for details.}
\label{fig:FidelityTransient}
\end{figure*}

In Fig.~\ref{fig:CPSlong}, we have found at best a CPS of $\sim\pi$ in correspondence with
fidelities ${\mathcal F}(t_{int})$ and ${\mathcal F}^c(t_{int})$ equal to $60\%$ and $80\%$,
respectively. This is due to the general presence of a \emph{trade-off between the size of the CPS and of the gate fidelity},
as well as to the atomic dephasing~\footnote{Without the dephasing, steady-state fidelities reach the values of $77\%$ and $83\%$
for unconditional and conditional cases, respectively}. This is an important
result of our paper, which actually holds true in \emph{any} EIT-based nonlinear optics systems. In fact, both the conditional and the unconditional gate fidelity approach $1$ in the limit of very small $g_j$, but this
limit yields a CPS which becomes appreciable only for unrealistically long gas cells. For the parameters of Fig~\ref{fig:CPSlong}, $v_g = 5.8\times 10^6$, which requires a cell of length $v_g \times t_{int} = 5.8\times 10^6 \textrm{ m/s} \times (500/\gamma) \simeq 81$ m.
Therefore a larger CPS requires a larger ratio $g_j\sqrt{N_a}/\Omega_j$.
This condition however increases the population of the collective atomic states $|e_1^{(n_p,n_t)}\rangle $ and $| e_5^{(n_p,n_t)}\rangle $
at the expense of the initial atomic state $\ket{3}$, thus unavoidably decreasing the gate fidelity. Similar conclusions hold
for other options, such as increased detunings $\delta_j$, or adjusting two-photon detunings $\epsilon_{ij}$. Therefore,
just the pure coherent unitary evolution of the system, governed by the Hamiltonian of Eq.~(\ref{eq:coll_ham}) causes this inherent trade-off.

\subsection{Short Interaction Time} \label{sec:SSShort}

To further illustrate our findings, we calculate the CPS and the gate fidelities in the range of parameters where the total interaction
time is an order of magnitude smaller than in Sec.~\ref{sec:SSLong}.
The CPS and the gate fidelities are calculated as described in Sec.~\ref{sec:SSLong}, and the results are shown in Fig.~\ref{fig:CPSshort}. To obtain a CPS of the order of $\pi$ in a shorter interaction time ($t_{int}\sim 50/\gamma$),
we have assumed a larger ratio $g_j\sqrt{N_a}/\Omega_j$. The trade-off between the amount of accumulated nonlinear phase shift
and the gate fidelity is now even more pronounced: we find at best a
CPS of $\sim\pi$ in correspondence with fidelities ${\mathcal F}(t_{int})$ and ${\mathcal F}^c(t_{int})$ equal to $65\%$ and $73\%$, respectively.
As expected, having a stronger atom-field coupling enhances the processes lowering the fidelity. The system ends up with a large CPS faster, but this is achieved with a final state in which the probability of loosing the probe and trigger photons
by spontaneous emission or within the atomic medium is no more negligible. We notice that in this short interaction time case
dephasing does not have an appreciable effect, that is, the results without dephasing are indistinguishable from those
with dephasing shown in Fig.~\ref{fig:CPSshort}, due to fact that dephasing rates in a dilute gas are typically much smaller than decay rates.
The only possible way to circumvent this trade-off is to explore the transient regime, which will be discussed in the following Section.

\section{QPG Operation in Transient Regime} \label{sec:transient}

In Sec.~\ref{sec:SSOperation}, we have found that the QPG operation of EIT-based nonlinear system in a steady-state is plagued by the trade-off
between the phase shift size and the gate fidelity. In an attempt to find favorable conditions for the QPG operation,
we consider the transient regime, when $\gamma t_{int}\lesssim 1$. As discussed above, in order to accumulate a significant CPS
in such a short time one has to consider the strong coupling regime with a large ratio $g_j\sqrt{N_a}/\Omega_j$.
Therefore, the trade-off between fidelity and a large nonlinear interaction is present also in the transient regime. However,
when $g_j\sqrt{N_a}/\Omega_j$ is large, the transient dynamics is characterized by
Rabi-like oscillations of the atomic populations and of the photon number, determining, as a consequence, coherent oscillations of the gate fidelity.
In such a case one cannot exclude the existence of special values of the interaction time $t_{int}$ corresponding to a maximum of the gate fidelity
close to one, and at the same time, to a value of the CPS close to $\pi$.

\begin{figure*}[ht]
\begin{center}
\includegraphics[scale=0.9]{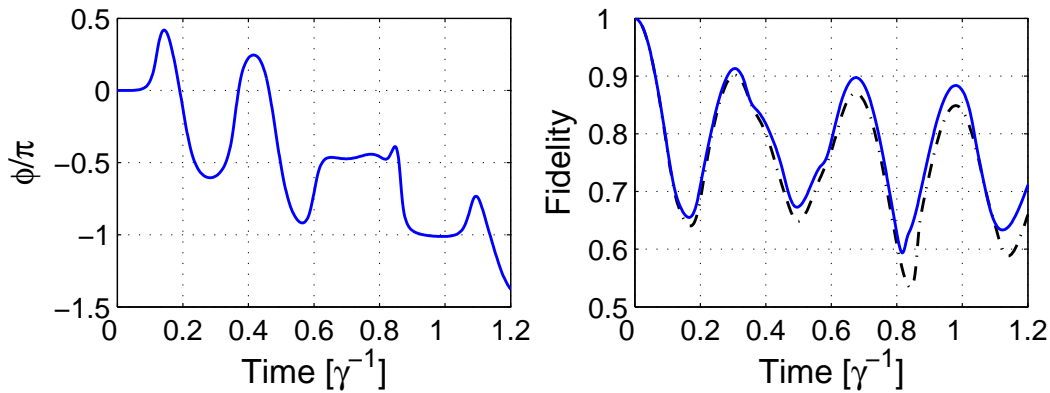}
\caption{Conditional phase shift (\textit{left}) and the fidelity (\textit{right}) of a QPG operation for $N_a = 10^8$, $\delta_2 = \delta_3 =
    6\gamma$, $\epsilon_{12} = \epsilon_{34} = 0.05\gamma$, $g_p = g_t = 0.0009\gamma$ and $\Omega_1 = \Omega_4 = \gamma$. We have taken equal decay rates, $\gamma_{21}=\gamma_{23}=\gamma_{25}=\gamma_{41}=\gamma_{43}=\gamma_{45}=\gamma/3$. For $^{87}$Rb, $\gamma =
    2\pi \times 6$ MHz, giving the interaction time (i.e. pulse length) of $\simeq 25$~ns for $|\phi| \simeq \pi$. \textit{Right}: solid line denotes
    the deterministic fidelity, while dot-dashed line denotes the conditional fidelity.}
\label{fig:CpsFtransient}
\end{center}
\end{figure*}

We show that this fact is actually possible in
Fig.~\ref{fig:FidelityTransient}, where we see that a CPS of $\sim \pi$ radians is obtained
in the transient regime for $t_{int}\approx 0.4/\gamma \sim 10$ ns. At the same interaction time, the unconditional
gate fidelity (Fig.~\ref{fig:FidelityTransient}, \textit{right}, full line) is about $94\%$, while the conditional gate fidelity reaches
the value of $99\%$ (Fig.~\ref{fig:FidelityTransient}, \textit{right}, dashed line). The conditional gate fidelity is obtained in correspondence
with a success probability of the gate equal to $0.94$, calculated from the norm of the Monte-Carlo wave function~\cite{Carmichael93}.
The probe and trigger group velocities are calculated to be $v_g \simeq 3 \times 10^6$ ms$^{-1}$, yielding a gas cell length
$L = v_g t_{int} \simeq 3.1$ cm. The value of $g_j$ yields an interaction volume $V \simeq 2 \cdot 10^{-3}$ cm$^3$,
corresponding to a gas cell diameter of about $330$ $\mu$m and to an atomic density $N_a/V \simeq 5 \cdot 10^{10}$ cm$^{-3}$.

To give a further example and to deepen our discussion of QPG performance and the CPS-fidelity trade-off, we now show the optimal results for the experimentally available pulses produced in the experiment of Darqui\'{e} \textit{et al.}~\cite{Darquie2005} (see Fig.~\ref{fig:CpsFtransient}). The length of a pulse produced in~\cite{Darquie2005} is 26 ns ($\simeq \gamma^{-1}$). The optimal parameters of Fig.~\ref{fig:CpsFtransient} give the unconditional gate fidelity (Fig.~\ref{fig:CpsFtransient} \textit{right}, solid line) of $85\%$, while the conditional fidelity reaches $89\%$ (Fig.~\ref{fig:CpsFtransient} \textit{right}, dot-dashed line). The probe and trigger group velocities are calculated to be $v_g \simeq 1.51 \times 10^6$ ms$^{-1}$, yielding a gas cell length $L = v_g t_{int} \simeq 3.82$ cm. The value of $g_j$ yields an interaction volume $V \simeq 10^{-1}$ cm$^3$, corresponding to a gas cell diameter of about $910$ $\mu$m and to an atomic density $N_a/V \simeq 10^{9}$ cm$^{-3}$. So, taking the longer wavepackets (i.e. longer interaction times) means that the CPS-fidelity trade-off becomes increasingly important, and the top value of fidelity decreases slightly with respect to its optimal value. 

Note how in both cases (Figs.~\ref{fig:FidelityTransient} and~\ref{fig:CpsFtransient}) high values of fidelity could be obtained for a values of CPS lower than $\pi$ radians. This implies the possibility of the implementation of a \textit{universal} quantum gate~\cite{Lloyd95}, which requires only $\phi \neq 0$.

A comment about this calculation of the common group velocity of the two wave-packets, $v_g$, is in order. As mentioned earlier, EIT is stationary phenomenon, and in fact, the conventional $v_g$ is a steady-state quantity which it is obtained from the susceptibility $\chi$ according to
\begin{equation}
v_g = c\left[1+\frac{1}{2} {\rm Re}[\chi] + \frac{\omega_0}{2} \left(\frac{\partial
{\rm Re}[\chi]}{\partial \omega} \right)_{\omega_0}\right]^{-1} \label{eq:ng}
\end{equation}
($\omega_0$ is the central frequency of wave-packet), where the susceptibility of the $j$-th field, $\chi_j$ ($j=p,t$), is evaluated from the
associated steady-state atomic coherence $\rho_j^{ss}$ as
\begin{equation}\label{eq:avchi2}
\chi_{j} = \frac{N |\mu_{j}|^2}{V\hbar \varepsilon_{0}\Omega_j}
\rho_j^{ss}.
\end{equation}
Instead, the above results are obtained in the transient regime where $\gamma t_{int} < 1$, and for this reason we have estimated the
group velocity in a different way. We have evaluated the relevant time-dependent atomic coherence $\rho_j(t)$
and the corresponding ``instantaneous susceptibility'' $\chi_j(t)$ from the reduced atomic density matrix $\rho_{red}(t)
= \textrm{Tr}_{fields} \{\rho (t)\}$, with $\rho (t)$ being the solution of Eq.~(\ref{eq:mastercoll2}).
The corresponding ``instantaneous''group velocity $v_g(t)$ has been then averaged over the time interval between $0$ and $t_{int}$,
providing in this way our estimate of the ``transient'' non-stationary group velocity of the single-photon wave-packets.
For the parameters of Fig.~\ref{fig:CpsFtransient}, this non-stationary
$v_g$ is approximately equal to $c/100$ and it is about one order of magnitude smaller than the conventional $v_g$ obtained from the
steady-state susceptibility. This appreciable slowing down of the group velocity is a signature of
a sort of ``non-stationary'' EIT process.

In order to verify that the non-stationary dynamics is really reminiscent of
EIT, in the next subsection we compare these results with a numerical study of the three-level ladder atomic scheme
(see Fig.~\ref{fig:3levelscheme}), yielding XPM without EIT.
Here we anticipate that we have found a smaller gate fidelity ($\sim 78\%$) for a corresponding set of parameters, providing therefore
further support to the presence of a moderate, non-stationary EIT
process in the transient dynamics of our five-level \emph{M} scheme. This is because in the ladder scheme, the increase in the size of the nonlinearity (and thus the CPS) is accompanied by an increase of losses, unlike the case of any EIT-based scheme~\cite{Schmidt96}. For this reason we also expect that our \emph{M} scheme outperforms ladder scheme in the case of fully optimized fidelity for $\pi$ phase shift.

\subsection{The Conventional Three-Level Scheme} \label{sec:ladder}

\begin{figure}[t]
\includegraphics[scale=0.8]{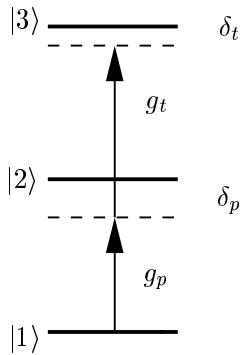}
    \caption{Energy levels of the ladder scheme. $g_{p,t}$ denote couplings of the quantized probe and trigger fields to their
    respective transitions. $\delta_{p,t}$ are detunings of the probe and trigger fields from resonance.} \label{fig:3levelscheme}
\end{figure}

The atomic ladder scheme (see Fig.~\ref{fig:3levelscheme}) is well-known to exhibit XPM of the two fields involved~\cite{Schmidt96}.
In order to achieve a reasonable size of cross-phase shift, the detuning of the intermediate state $\delta_p$ needs to be large. This
minimizes spontaneous emission $(\sim \gamma_2/\delta_p^2)$, but also the size of XPM $\sim 1/\delta_p^2$.

In order to evaluate the XPM in a manner comparable to what we have done for the \textit{M}-scheme, we make similar assumptions and arrive at a description analogous to the one described in Sec.~\ref{sec:model}. The Hamiltonian is now given by
\begin{eqnarray}\label{eq:ladder_ham}
H_3 &=& \hbar\delta_{p}\hat{S}_{22} + \hbar (\delta_p-\delta_t)\hat{S}_{33} + \hbar g_{p}\sqrt{N_{a}} \left(\hat{a}_{p}\hat{S}_{21} + \hat{S}_{12}\hat{a}_{p}^{\dagger}
\right) \nonumber \\
&&+ \hbar g_{t} \sqrt{N_{a}} \left(\hat{a}_{t}\hat{S}_{23} + \hat{S}_{32}\hat{a}_{t}^{\dagger} \right).
\end{eqnarray}
Following the same reasoning as in Sec.~\ref{sec:model}, we arrive at the effective master equation (we neglect here atomic dephasing)
\begin{eqnarray}
\dot{\rho} &=& {\mathcal L}_3\rho = -\frac{i}{\hbar}\left[ H_3,\, \rho\right] \nonumber \\
&& + \frac{\gamma_{21}}{2}\left( 2\hat{S}_{12}\rho\hat{S}_{21}
- \hat{S}_{21}\hat{S}_{12}\rho - \rho\hat{S}_{21}\hat{S}_{12} \right) \nonumber \\
&& + \frac{\gamma_{32}}{2}\left( 2\hat{S}_{23}\rho\hat{S}_{32}
- \hat{S}_{23}\hat{S}_{32}\rho - \rho\hat{S}_{23}\hat{S}_{32} \right),
\label{eq:masterladder}
\end{eqnarray}
where $\gamma_{12}$ and $\gamma_{23}$ denote the spontaneous emission rates from levels $\ket{2} \mbox{and} \ket{3}$ to levels $\ket{1} \mbox{and} \ket{2}$
respectively, and the operators $\hat{S}_{ij}$ denote collective atomic operators, in the spirit of Eq.~(\ref{eq:collective1}).

\begin{figure}[t]
\includegraphics[scale=0.8]{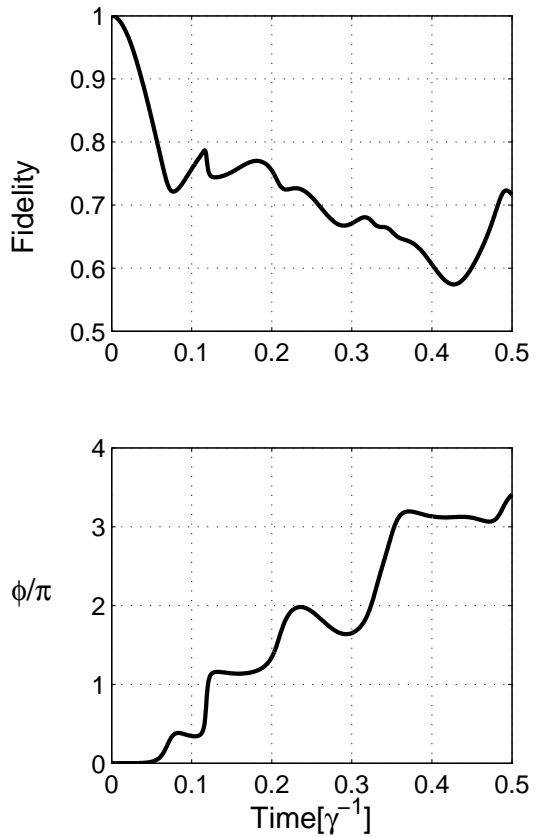}
    \caption{Average fidelity (top figure) and conditional phase shift (bottom figure) as a function of time for the three-level
    ladder scheme of Fig.~\protect\ref{fig:3levelscheme} and for $N_a=10^8$, $\delta_p=10\gamma$,
    $\delta_t=0$ and $g_p=g_t=0.0022\gamma$. The spontaneous emission rate is $\gamma_{21} = \gamma_{32} = \gamma = 2\pi\times 6$ MHz.}
\label{fig:Ladder}
\end{figure}

The results of the calculation of unconditional quantities are shown in Fig.~\ref{fig:Ladder}, for a parameter regime comparable to that discussed in relation to Figs.~\ref{fig:FidelityTransient} (see also Ref.~\cite{rapcomm}) for the five-level \emph{M} scheme. The atoms are assumed to be in the collective state $\bigotimes_{i=1}^{N_a} \ket{2}_i$ initially, as this is found to give better results. The reason is a more efficient photon-photon interaction since the initial state is symmetric with respect to probe and trigger photons. At the interaction time $t_{int} \approx 0.12/\gamma$, the CPS reaches the value $\sim \pi$ and at the same time the unconditional fidelity is found to be $\sim 78\%$. It is possible to see that the conditional fidelity, even though higher, always remains significantly lower than that obtained in the \emph{M}-scheme.

Therefore, we conclude that the optimal results for the QPG operation can be found in the ``transient EIT-regime''.
The general trade-off between the nonlinear phase shift and the fidelity is still present, but it is compensated by the transient oscillations
in the populations of atomic levels. In fact, the numerical results
show that, in the parameter regime under consideration,
the population of the excited states $|e_2^{(n_p,n_t)}\rangle $ and $| e_4^{(n_p,n_t)}\rangle $ is always negligible, and one has coherent
oscillations of the population between the states $|e_1^{(n_p,n_t)}\rangle $, $\bigotimes_i \ket{3}_i \otimes |n_p,n_t\rangle $ and $| e_5^{(n_p,n_t)}\rangle $.
At the interaction time $t_{int}$ corresponding to the maxima of the gate fidelity in Fig.~\ref{fig:CpsFtransient}, atoms are largely found in state
$\bigotimes_i \ket{3}_i \otimes \ket{n_p,n_t}$, and the relative populations \emph{and} phase relations between the states of the two photonic qubits
are consistent with the ``ideal'' state of Eq.~(\ref{eq:psi_id}).

\section{Experimental verification of the QPG operation} \label{sec:exper}

In this Section, two possible schemes for experimental implementation are discussed. First is the detection in the occupation number logical basis, and the second is the detection in the polarization logical basis. The two bases are identical from the point of view of theoretical treatment, however, their implementation is different in practice.

\begin{figure}[h]
\includegraphics[angle=0,width=.45\textwidth]{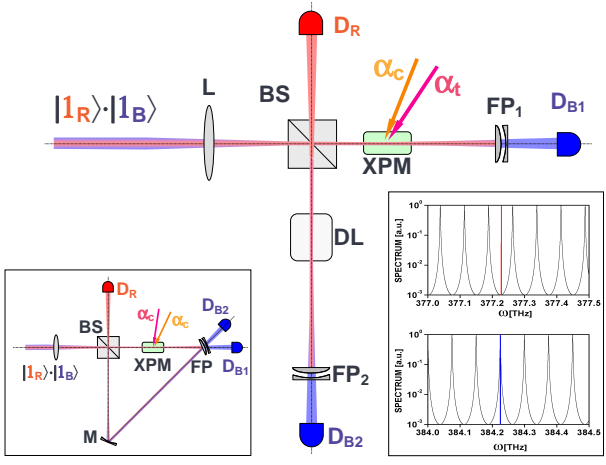}
\caption{\label{fig:Fig_EIT} (Color online) Scheme of the proposed experiment for
a measurement of the non-linear phase shift in a QPG. A
Michelson-like interferometer with a two-photons input state
$|1_{\rm R}\rangle|1_{\rm B}\rangle$, {\it probe} and {\it
trigger}, respectively, allows to measure the non-linear phase
induced by the XPM on the logical basis of the qubits, which
coincide with the two lowest Fock states. Two intense classical
fields $\alpha_t$ ({\it tuner} with Rabi frequency $\Omega_1$) 
and $\alpha_c$ ({\it coupler}, with Rabi frequency $\Omega_4$), 
are necessary to the 5-level XPM process. L is a
lens for the mode matching in the EIT medium. BS a 50/50 beam splitter.
DL a delay line. ${\rm FP}_{1,2}$ Fabry-Perot cavities. ${\rm
D}_{R,B1,B2}$ are avalanche-photodiodes APDs. {\it Left Inset --}
Scheme using a Sagnac interferometer for avoiding the optical path
difference. M is a mirror. {\it Right Inset --} Frequency spectrum [arb. units]
for a 2~mm FP cavity length and finesse equal to $10^3$. The spectra 
of the probe and trigger photon are also shown in the plots as lines.}
\end{figure}

{\it Occupation number logical basis --} In this section we
describe a Michelson-like interferometer (see
Fig~\ref{fig:Fig_EIT}) for two-photons product state $|1_{\rm
R}\rangle|1_{\rm B}\rangle$ with the `right' $\sigma^{-}$ circular
polarization, where R (red) refers to the probe field and B
(blue) to the trigger. We show that the interferometer is
able to reveal and measure the QPG phase shift. The probe and
trigger fields with a bandwidth of $40\div100$~MHz (corresponding
to 25$\div$10~ns 1/e half width pulse duration~\cite{Darquie2005}) are separated in
frequency by $\sim$7~THz and are resonant with the $^{87}{\rm Rb}$
hyperfine transitions $D_1F=2\rightarrow F^\prime =1$  at 794.7~nm
(377.228~THz), and $D_2F=2\rightarrow F^\prime =1$ at 780.2~nm
(384.225~THz), respectively.

The interferometer is realized with the help of a 50/50 beam splitter (BS),
using a Fabry-Perot cavities (${\rm FP_{1,2}}$) instead of
mirrors. The FPs reflect back the probe field, which is then
superimposed on the BS and detected by an APD (${\rm D_{R}}$), and
transmit the trigger field detected by an APD in each arm (${\rm
D_{B1}}$ e ${\rm D_{B2}}$). This implies that only the trigger
frequency is resonant with the FPs' cavity, which has a cavity
length of 2~mm corresponding to a FSR of 74.85~GHz, while the
probe frequency falls in the middle of the previous 93$^{\rm th}$
and 94$^{\rm th}$ FSRs. According to the photon bandwidth, a
finesse of $10^3$ determines a reflectivity for the probe field of
99.9$\%$ (see the spectra in the {\it Right Inset} of
Fig.~\ref{fig:Fig_EIT}).

Since the frequency-bandwidths of the two photons are well distinguished 
and the FPs filter out the trigger field, this apparatus determines
an interferometer for the probe field only. The coincidence
probabilities $P({\rm R,B1})$, between ${\rm D_{R}}$ and ${\rm
D_{B1}}$, and $P({\rm R,B2})$, between ${\rm D_{R}}$ and ${\rm
D_{B2}}$, post-select the events in which the trigger photon is. 
In this case the coincidence probabilities are
equal and given by
\begin{eqnarray}\label{Coinc_Prob_NoQPG}
  P({\rm R,B1}) = P({\rm R,B2}) = [1 +\cos\Phi]/8
\end{eqnarray}
where $\Phi$ represents the phase difference due to the different
optical paths of the two arms experienced by the probe. In arm 2 a
delay line (DL) is added to compensate the difference in the
optical path and to scan for the interference pattern. In the {\it
Left Inset} of Fig.~\ref{fig:Fig_EIT} a Sagnac-like
version of the interferometer is shown, which allows for an
auto-compensation of the optical path delay as the two arms
coincide.

\begin{figure}[h]
\includegraphics[angle=0,width=.45\textwidth]{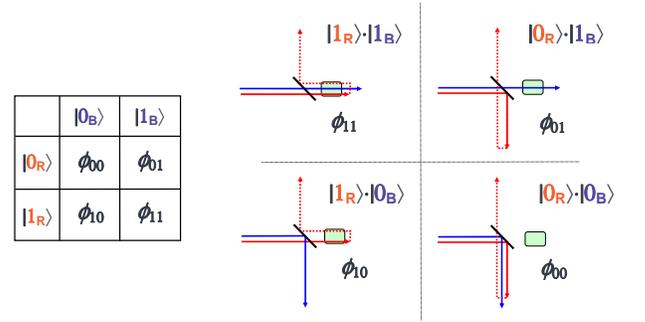}
\caption{\label{fig:Fig_Table} (Color online) Truth-table of the QPG for a logical
basis of the qubits determined by the two lowest Fock states. On
the right, diagrams corresponding to the probability amplitudes
after the action of the BS and the FPs on the two-photons input
state are shown.}
\end{figure}

When an EIT--based XPM system is considered in one arm, say arm 1,
a non-linear contribution to the phase is added by the QPG,
whether the trigger photon is present in the arm 1 or not. The XPM
requires two intense classical fields resonant to the
$D_2F=1\rightarrow F^\prime =1$ (384.232~THz) and
$D_1F=1\rightarrow F^\prime =1$ (377.235~THz), $\sigma^{-}$
circularly polarized tuner field ($\alpha_t,\ \Omega_1$), and
$\sigma^{+}$ circularly polarized coupling field
($\alpha_c,\ \Omega_4$), respectively. The coincidence probabilities $P({\rm
R,B1})$, between ${\rm D_{R}}$ and ${\rm D_{B1}}$, and $P({\rm
R,B2})$, between ${\rm D_{R}}$ and ${\rm D_{B2}}$, enable post-selection of the
events in which the trigger photon is in the arm 2 and hence 
the interference pattern is given by Eq.~(\ref{Coinc_Prob_NoQPG}).
Instead a non-linear contribution to the phase is expected when
detections on ${\rm D_{R}}$ are post selected by the detection on
${\rm D_{B2}}$. Fig.~\ref{fig:Fig_Table} represents the diagram of
the four amplitude probabilities after the action of the BS and
the FPs on the two-photons state. The non-linear phase $\phi_{11}$
is added only to the first diagram.
\\
According to the truth-table and the amplitudes of
Fig.~\ref{fig:Fig_Table}, the coincidence probabilities can be
evaluated to be
\begin{eqnarray}\label{Coinc_Prob_QPG}
  P({\rm R,B1}) &=& [1 +\cos(\widetilde{\Phi} + \phi)]/8 \\
  P({\rm R,B2}) &=& [1 +\cos\widetilde{\Phi}]/8
\end{eqnarray}
with $\tilde{\Phi} = \Phi + (\phi_{10}-2\phi_{00} + \phi_{+0})$.
$\phi_{+0}$ is the phase due to the probe photon reflected in arm 1
with `wrong' $\sigma^{+}$ circular polarization. The phases
$\phi_{10}$ and $\phi_{00}$ have been introduced in Eq.~(\ref{eq:def_cps}) as well as the
conditional phase shift $\phi=\phi_{11}-\phi_{01}-\phi_{10}+\phi_{00}$ of the QPG. The
phase difference between the two interference patters determined
by the two coincidence probabilities determines univocally the
value of $\phi$. In the case of an ideal QPG for which $\phi =
\pi$ the two coincidence probabilities have opposite phases.

{\it Polarization logical basis --} The previous proposal
indirectly tests the QPG based on the XPM detecting the non-linear
phase shift by a Michelson interferometer and coincidence
measurements. A direct measurement of the truth table or a test on
a general qubit state requires a control and measurement of a
superposition of vacuum and one photon state. While the generation
of a superposition of vacuum and single photon state has been
already achieved~\cite{Darquie2005}, the measurement of such a
superposition requires also a homodyne
measurement~\cite{Babichev04,Zavatta04}. However a logical basis
for the qubits can be chosen as the circular polarization basis of
the probe and trigger photons. A test of the QPG will then require
the detection of both photons thus avoiding a problematic production and 
detection of Fock state superpositions.

The experimental setup is the same as in Fig.~\ref{fig:Fig_EIT},
but the qubits are now encoded in the polarization of the input
two-photon state.
\begin{figure}[h]
\includegraphics[angle=0,width=.45\textwidth]{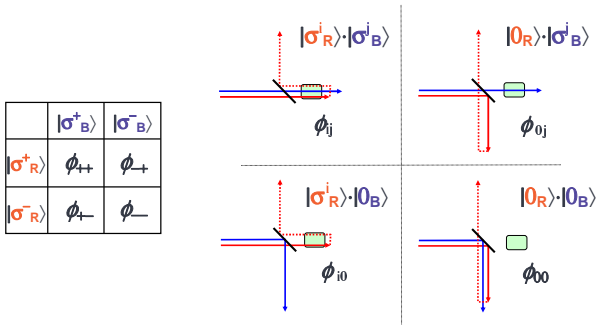}
\caption{\label{fig:Fig_EIT_Pol} (Color online) Truth-table of the QPG for a
logical basis of the qubits determined by orthogonal circular
polarization basis. On the right the diagrams corresponding to the
probability amplitudes after the action of the BS and the FPs on
the two-photons input state.}
\end{figure}
According to the truth-table and the amplitudes in
Fig.~\ref{fig:Fig_EIT_Pol} it is possible to derive the
coincidence probabilities $P({\rm R_i,B1_j})$ and $P({\rm
R_i,B2_j})$, with $i,j=\{+,-\}$, as
\begin{eqnarray}\label{Coinc_Prob_QPG_pol}
  P({\rm R_i,B1_j}) &=& [1 +\cos(\overline{\Phi} + \overline{\phi}_{ij})]/8 \\
  P({\rm R_i,B2_j}) &=& [1 +\cos\overline{\Phi}]/8
\end{eqnarray}
with $\overline{\Phi} = \Phi + (\phi_{i0} -2\phi_{00} +
\phi_{(i\oplus1)0})$, where $i\oplus1$ is the sum mod~2. The phase
$\overline{\phi}$ is now given as $\overline{\phi}_{ij} =
\phi_{ij}-\phi_{i0}-\phi_{0j}+\phi_{00}$, where $\phi_{i0}$ is the
phase due to the EIT for the probe photon with polarization $i$
and no trigger photon present, and same meaning for the other
phases. Note that for $i = j = `-' \equiv `1'$ one obtains
the previous expression for the QPG phase in the logical Fock-state 
basis. Four possible choices of the probe and trigger
polarizations determine the phases
\begin{eqnarray}\label{Phase_Pol_Measur}
\begin{tabular}{c}
    $\overline{\phi}_{--} = \phi_{--}-\phi_{-0}-\phi_{0-}+\phi_{00}$\,
        \\
    $\overline{\phi}_{-+} = \phi_{-+}-\phi_{-0}-\phi_{0+}+\phi_{00}$\,
        \\
    $\overline{\phi}_{+-} = \phi_{+-}-\phi_{+0}-\phi_{0-}+\phi_{00}$\,
        \\
    $\overline{\phi}_{++} = \phi_{++}-\phi_{+0}-\phi_{0+}+\phi_{00}$\,,
\end{tabular}\end{eqnarray}
which satisfies the relation $\overline{\phi}_{--}
-\overline{\phi}_{+-} -\overline{\phi}_{-+} + \overline{\phi}_{++}
= \phi_{--} -\phi_{+-} -\phi_{-+} + \phi_{++} = \phi$. In the case
of an ideal EIT
%(specify What/How)
for which
$\phi_{+0}=\phi_{0+}=\phi_{++}=\phi_{00}$,
$\phi_{-+}=\phi_{-0}=\phi_{R}$ and $\phi_{+-}=\phi_{0-}=\phi_{B}$,
where $\phi_{A,B}$ are the phases acquired by the single photons,
we have $\overline{\phi}_{--} = \phi$, and $\overline{\phi}_{-+}
=\overline{\phi}_{+-} =\overline{\phi}_{++} = 0$. In a way the
phases between the two coincidence interference patterns allow a
measurement of the QPG phases in the diagonal basis, i.e. in the
single qubit states for which the only non zero phase shift is the
conditional phase shift $\phi$ of the QPG.

\begin{figure}[h]
\includegraphics[angle=0,width=.45\textwidth]{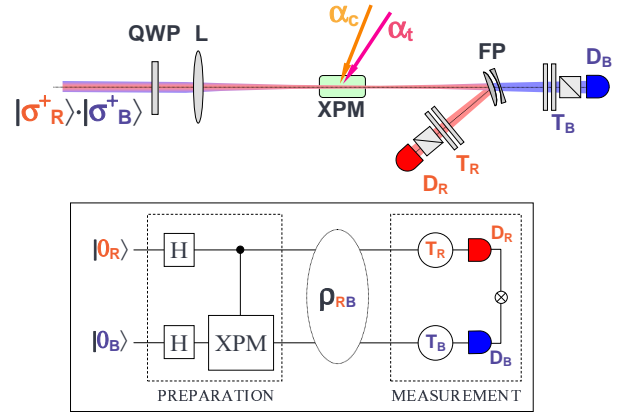}
\caption{\label{fig:Fig_EIT_Bell} (Color online) Scheme of the proposed
experiment for a complete characterization of the QPG. Two photons
in the $\sigma^{+}$ polarization state (logical state
$|0_{R}\rangle\otimes|0_{B}\rangle$) are transformed by a QWP,
corresponding to two Hadamard single-qubit gates, and then to the
XPM medium (QPG). A Fabry-Perot cavity with the same parameter as before
transmits the trigger photon and reflect the probe to two
tomographic measurement systems (${\rm T_{R,B}}$) and detected by
APDs.}
\end{figure}

{\it General polarization qubit input states --} The polarization
logical basis allows for a direct observation of coherence and
production of entanglement as necessary conditions for a QPG. 
Assume that the information is encoded in the polarization state of
two photons, and then sent into the EIT-based XPM system for QPG, as
shown in Fig.~\ref{fig:Fig_EIT_Bell}. The output photons, as
shown in Fig.~\ref{fig:Fig_EIT_Bell}, are split by a dichroic
mirror (a tilted FP cavity with the same parameters as before)
and then collected in two APDs (${\rm D}_{R}$ and ${\rm D}_{B}$)
for coincidence counting. In front of each detector a tomographic
system~\cite{James01} constituted by a QWP, HWP and a PBS, is placed for the
complete reconstruction of the polarization state of the output,
thus providing the information on coherence properties of the gate.
It has also been shown~\cite{turch} that an input state for
the QPG given by $[(|\sigma^{+}_{R}\rangle +
|\sigma^{-}_{R}\rangle)
    \otimes
  (|\sigma^{+}_{B}\rangle + |\sigma^{-}_{B})\rangle)]/2$
can quantify the entanglement of the output state, for which the
CHSH inequality is $2\sqrt{1 + \sin^2\phi}$ where 2 is the upper
classical limit.

\section{Conclusion} \label{sec:conclusion}

In conclusion, our study shows that the implementation of efficient EIT-based nonlinear two-qubit gates for travelling single-photons is possible,
even though experimentally challenging. The main limitation is due to the existence of a trade-off between the size of the CPS and the
fidelity of the gate, limiting the achievable gate fidelity in the stationary regime,
but which can be partially bypassed in the transient regime. Since this trade-off is a general consequence of the coherent
interaction between the atomic medium and the single-photon wave-packets, we expect that these considerations apply to all EIT-based
crossed-Kerr schemes \cite{Lukin00,Ottaviani03}, regardless the specific level scheme considered. Instead, this consideration
does not apply to situations where the nonlinearity comes from independent processes (e.g. collisions of dipole-dipole interactions)~\cite{Masalas04},
nor the similar solid-state based processes~\cite{Longdell04}.

\section{Acknowledgements}
This work has been partly supported by the European Commission through FP6/2002/IST/FETPI
SCALA: 'Scalable Quantum Computing with Light and Atoms', Contract No 015714, CONQUEST network, MRTN-CT-2003-505089 and under the Integrated Project Qubit Applications (QAP) funded by the IST directorate, Contract No 015848. GDG also acknowledges financial support from the Ministero della Istruzione, dell'Universita' e della Ricerca (PRIN-2005024254 and FIRB-RBAU01L5AZ).

\end{document}